\documentclass{article}
\usepackage{epsfig,amssymb,amsmath}
\renewcommand{\arraystretch}{1.5}

\begin{document} 
 
\title{Approximation schemes for the dynamics of diluted spin models:
the Ising ferromagnet on a Bethe lattice} 
\vskip 10pt 
\author{ 
Guilhem Semerjian$^{1}$ and Martin Weigt$^{2}$
\\ 
$^1$Laboratoire de Physique Th{\'e}orique de l'{\'E}cole Normale  
Sup{\'e}rieure,  
\\ 
24 rue Lhomond, 75231 Paris Cedex 05, France 
\\ 
$^2$Institut f\"ur Theoretische Physik, Universit\"at G\"ottingen,
\\
Tammannstr. 1, 37077 G\"ottingen, Germany
\\
{\tt guilhem@lpt.ens.fr, weigt@theorie.physik.uni-goettingen.de}
} 
\maketitle 

\begin{abstract} 
  We discuss analytical approximation schemes for the dynamics of
  diluted spin models. The original dynamics of the complete set of
  degrees of freedom is replaced by a hierarchy of equations including
  an increasing number of global observables, which can be closed
  approximately at different levels of the hierarchy. We illustrate
  this method on the simple example of the Ising ferromagnet on a
  Bethe lattice, investigating the first three possible closures,
  which are all exact in the long time limit, and which yield more and
  more accurate predictions for the finite-time behavior. We also
  investigate the critical region around the phase transition, and the
  behavior of two-time correlation functions.  We finally underline
  the close relationship between this approach and the dynamical
  replica theory under the assumption of replica symmetry.

\vspace{.5cm} 

\noindent LPT-ENS 04/04
\end{abstract} 
 
\section{Introduction}
The last few years have seen a considerable increase in the research
activity concerning diluted and disordered spin models. This recent
interest is mainly driven by two very different, but technically
closely related motivations.

The first one aims at understanding common fundamental features of
glassy systems, or more generally of out-of-equilibrium problems.
Besides various other approaches \cite{Go,RiSo}, the study of
disordered spin models has led to important insights and given rise to
the use of general concepts like, e.g., replica symmetry breaking on
the static side \cite{Beyond}, or effective temperatures on the
dynamic one \cite{BoCuKuMe,Cu}. Most of these studies were performed
on mean-field models, and the validity of these concepts for
finite-dimensional systems is still under discussion.  For spin
models, mean-field has long been another name for fully-connected,
i.e.~models where each degree of freedom interacts with each other.
More recently, a lot of effort was invested in the study of diluted
models \cite{Mo,MePa,FrMeRiWeZe,BiMe,WeHa,PiTaCaCo,RiBiMaMe}. These
are still mean-field like in the sense that they do not have an
underlying geometric structure, and are thus analytically easier to
treat than finite-dimensional problems. At the same time, they have
finite connectivity: each degree of freedom interacts only with a
finite number of neighbors, and the concept of a local environment is
well-defined. This allows, e.g., to introduce microscopically
motivated interactions. Diluted models can therefore be considered as
an intermediate step between the fully-connected and the
finite-dimensional case.

A second major motivation for the study of diluted spin models arises
from their close connection with a very interesting class of
combinatorial optimization problems \cite{GaJo}, including famous
problems like satisfiability of Boolean formulas and graph coloring
for instance.  These problems show phase transitions in the
statistical properties of their solutions as well as in the dynamical
behavior of algorithms \cite{AI,TCS,PTAC}, and have thus been extensively
studied using tools and concepts originally devised for statistical
mechanics purpose \cite{MoZe,nature,BiMoWe,RiWeZe,MePaZe,MuPaWeZe}.

Despite the large interest in diluted models, many basic questions are
still open. Whereas the main technical obstacles in analyzing the
static behavior are solved by now, and the equilibrium properties of
these models are quite well-understood, the knowledge on the
dynamical side \cite{BaZe,MoRi,CuSe,SeCuMo} is still relatively poor.
Here we reinvestigate some approximative methods recently introduced
for analyzing the behavior of stochastic local search optimization
algorithms \cite{SeMo,BaHaWe}. We apply and generalize these ideas to
a purely physical model, in order to better understand the assumptions
behind the approximations made, and to assess their quality. We
concentrate on a very simple system, namely a ferromagnetic Ising
model defined on a diluted network of fixed connectivity, or Bethe
lattice. This example allows to easily present, test and understand
approximate methods which should be useful to treat more
interesting, glassy problems. In fact the non-equilibrium flavor of
this work comes from the study of transient relaxation from an
arbitrary initial configuration to thermal equilibrium, and not from
the absence of thermal equilibrium as in the case of glassy systems,
or from the lack of detailed balance as in previous investigations of
algorithms.

The paper is organized as follows. After this general introduction,
the model and its equilibrium behavior are presented in
Sec.~\ref{sec:model}. In Sec.~\ref{sec:dyn} the dynamical rules of the
model are presented, and a hierarchical approximation scheme is
developed. The first three approximations are compared to numerical
simulations in Sec.~\ref{sec:mc}, and they are further exploited
analytically to describe the critical behavior in
Sec.~\ref{sec:critical}. Sec.~\ref{sec:twotime} is dedicated to an
extension of the previous approximations to the analysis of two-time
quantities. In Sec.~\ref{sec:drt} we unveil the relation of our
approach to the dynamical replica theory proposed by Coolen and
Sherrington \cite{CoSh,LaCoSh}.  Finally, the results are summarized
in the last section, and possible future directions of research are presented.

\section{The model and its equilibrium behavior}
\label{sec:model}
We consider a ferromagnetic Ising model on a Bethe lattice, given by
its Hamiltonian
\begin{equation}
  \label{eq:Hamiltonian}
  H = -\frac 12 \sum_{i<j} J_{ij} (\sigma_i \sigma_j-1) 
\end{equation}
depending on the microscopic configuration $\vec \sigma =
(\sigma_1,...,\sigma_N)$ of the $N$ Ising variables $\sigma_i=\pm1,\ 
i=1,...,N$.  The couplings $J_{ij}$ take the value $+1$ whenever two
spin $\sigma_i$ and $\sigma_j$ are connected, and zero otherwise.
Compared to the usual form we have shifted the Hamiltonian by its
ground state value, and divided it by two. In this way, the
Hamiltonian simply counts the {\it number of unsatisfied edges}, i.e.
the number of edges carrying anti-parallel spins on their extremities.
It can thus be rewritten as
\begin{equation}
  \label{eq:Hamiltonian2}
  H = \sum_{i<j} J_{ij} \delta_{\sigma_i,-\sigma_j} \ .
\end{equation}
The reason why we choose this slightly modified form will become clear
below, it allows a simpler presentation of the dynamical equations.

As already said above, we consider this Ising model on a Bethe
lattice. In accordance with the recent use in statistical mechanics
\cite{MePa}, we define these as random regular graphs, i.e.~all
vertices have the same degree (number of neighboring sites) \cite{MoRe,Ne}. 
These graphs are locally
tree-like. In contrast to Cayley trees they have {\it no boundary},
all vertices have exactly $K$ neighbors and are thus equivalent. The
graph therefore contains loops for any $K>1$.  These are, however, of
length ${\cal O}(\ln N)$, i.e. they become long in the thermodynamic
limit $N\to \infty$ (with $K$ kept constant).  Note that the
randomness of these graphs appears only through these loops.  On
finite length scales they appear homogeneous due to their constant
vertex degree.

Before investigating the non-equilibrium dynamics of this model, we
shortly review its static behavior which can be easily solved using
the Bethe-Peierls iterative approach \cite{Ba,MePa}. For doing so, we
assume for a moment that our model is defined on a tree, the long
loops will be taken into account later as self-consistency
conditions.

Consider a given bond $(i,j)$, i.e. $J_{ij}=1$. Let us denote
$Z_{i|j}(\sigma_i)$ the partition function of the subtree rooted in
$i$, with $(i,j)$ deleted, and with a fixed value of spin $\sigma_i$.
These partition functions can be easily computed iteratively,
\begin{equation}
  \label{eq:iteration}
  Z_{i|j}(\sigma_i) = \prod_{k\neq j| J_{ik}=1} \left( 
    \sum_{\sigma_k=\pm1} Z_{k|i}(\sigma_k)\ \exp\{ -\beta\
    \delta_{\sigma_i,-\sigma_k} \} \right)\ ,
\end{equation}
where $\beta$ denotes the inverse temperature. Defining the cavity
field
\begin{equation}
  \label{eq:cavityfield}
  h_{i|j} = \frac 1{2\beta} \ln\left( \frac{Z_{i|j}(+1)}{Z_{i|j}(-1)}
  \right)\ ,
\end{equation}
we obtain the iterative equations
\begin{equation}
  \label{eq:fielditeration}
  2\beta h_{i|j} = \sum_{k\neq j| J_{ik}=1} \ln\left( 
    \frac{e^{\beta h_{k|i}}+e^{-\beta(1+h_{k|i})}}
    {e^{\beta(-1+h_{k|i})}+e^{-\beta h_{k|i}}}
  \right)\ .
\end{equation}
At this point, we take into account the fact that the model is not defined on a
tree, but that all vertices have the same vertex degree and are thus
equivalent. We are therefore looking for a self-consistent {\it
  homogeneous} solution of Eq. (\ref{eq:fielditeration})
with $h=h_{i|j}$ for all bonds $(i,j)$,
\begin{equation}
  \label{eq:field}
  h = \frac{K-1}{2\beta} \ln\left( 
    \frac{e^{\beta h}+e^{-\beta(1+h)}}{e^{\beta(-1+h)}+e^{-\beta h}}
  \right)\ .
\end{equation}
This equation has the obvious paramagnetic solution $h=0$. In fact,
at high temperature this solution is the only one. There appears,
however, a ferromagnetic phase transition at the critical inverse
temperature $\beta_c = \ln( K/(K-2))$. For lower temperatures, the
trivial solution is thermodynamically unstable, and two equivalent
ferromagnetic solutions $\pm h$ describe the equilibrium behavior of
the model.

Here and in the following, without any loss of generality, we
concentrate only on non-negative $h$.  Once its value is known, we can
immediately compute the Bethe free-energy density (total free-energy
divided by $N$),
\begin{equation}
  \label{eq:freeenergy}
  \beta f = (K-1)\ \ln \left( 2\cosh\left( \beta \frac K{K-1} h
    \right) \right)
-\frac K2\ \ln \left( 2\cosh(2\beta h) + 2e^{-\beta}\right)\ .
\end{equation}
Other interesting observables are the energy density 
\begin{equation}
  \label{eq:energy}
  e= \frac{\partial \beta f}{\partial \beta} 
  = Kh \tanh\left( \beta \frac K{K-1} h \right) - \frac K2\  \frac
  {2h\sinh(2\beta h) -e^{-\beta}}{\cosh(2\beta h) + e^{-\beta}}
\end{equation}
which equals $\frac K2 (1+e^\beta)^{-1}$ in the paramagnetic phase,
and the magnetization
\begin{equation}
  \label{eq:mag}
  m =  \tanh\left( \beta \frac K{K-1} h \right) \ ,
\end{equation}
which becomes positive for positive cavity fields $h$. Note that the
magnetization $m$ depends on the cavity field $h$ via the ``true''
effective field $Kh/(K-1)$ which incorporates the contributions of all
$K$ neighbors of a spin, cf. Eqs.
(\ref{eq:fielditeration},\ref{eq:field}).

To explain some steps of the dynamical approach, we also need the
probability $p_\sigma(u)$ that a randomly selected vertex has spin
value $\sigma$ and belongs to exactly $u$ unsatisfied edges, i.e. $u$
out of his $K$ neighbors have spin $-\sigma$. This quantity is given
at equilibrium by
\begin{eqnarray}
  \label{eq:p_sigma_u}
  p_+(u) &=& \frac 1{{\cal N}_v} {K \choose u} 
  e^{-\beta (2h+1) u} \ , \nonumber\\
  p_-(u) &=& \frac 1{{\cal N}_v} {K \choose u} 
  e^{-\beta u -2\beta h (K-u)}\ ,
\end{eqnarray}
with ${{\cal N}_v}$ being a normalization constant enforcing 
$\sum_{u,\sigma} p_{\sigma}(u)=1$. This can be
rewritten in terms of the energy density and the magnetization as
\begin{equation}
  \label{eq:p_sigma_u2}
  p_{\sigma}(u)=\frac{1+\sigma m}{2} {K \choose u} \left(\frac{2e}
    {(1+\sigma m)K}\right)^u \left(1-\frac{2e}{(1+\sigma
      m)K}\right)^{K-u} \ . 
\end{equation}
Similar expressions can be easily derived for the probability
$p_{\sigma_1 \sigma_2} (u_1, u_2)$ that a randomly selected edge has a
first (resp. second) end vertex of spin $\sigma_1$ (resp. $\sigma_2$),
and this vertex belongs to $u_1$ (resp. $u_2$) unsatisfied edges,
\begin{eqnarray}
  \label{eq:link_equ}
  p_{++} (u_1, u_2) &=& \frac 1{{\cal N}_e} {K-1 \choose u_1} {K-1 \choose
    u_2} e^{ -\beta(2h+1)(u_1+u_2) } \nonumber\\
  p_{+-} (u_1, u_2) &=& \frac 1{{\cal N}_e} {K-1 \choose u_1-1} {K-1 \choose
    u_2-1} e^{ -\beta(u_1+u_2-1)-2\beta h (u_1-1+K-u_2) } \nonumber\\
  p_{-+} (u_1, u_2) &=& \frac 1{{\cal N}_e} {K-1 \choose u_1-1} {K-1 \choose
    u_2-1} e^{ -\beta(u_1+u_2-1)-2\beta h (K-u_1+u_2-1) } \nonumber\\
  p_{--} (u_1, u_2) &=& \frac 1{{\cal N}_e} {K-1 \choose u_1} {K-1 \choose
    u_2} e^{ -\beta(u_1+u_2)-2\beta h(2K-u_1-u_2-2) } \ ,
\end{eqnarray}
where ${{\cal N}_e}$ is again determined by normalization.

\section{Dynamical approximation schemes}
\label{sec:dyn}

\subsection{Definitions}
\label{sec:defs}

We will study the following local stochastic dynamics of the model: in
each algorithmic step $T \to T+1$, a site $i$ is chosen at random,
$i\in\{1,...,N\}$. This site is characterized by its spin $\sigma_i$
and the number $u_i(\vec \sigma)=\sum_j J_{ij}
\delta_{\sigma_i,-\sigma_j}$ of its unsatisfied incident edges. The
spin is flipped to $-\sigma_i$ with probability
$W(u_i(\vec\sigma),\beta)$. We denote the new configuration, with spin
$\sigma_i$ flipped, by $F_i\vec\sigma$.

Obviously all edges incident to site $i$ which were unsatisfied become
satisfied, and vice versa. Hence the variation of the energy if the spin
is flipped equals $\Delta E= K-2u_i$. In order to reach thermal equilibrium at
inverse temperature $\beta$ in the long-time limit, we impose the
detailed balance condition under the form
\begin{equation}
  \label{eq:detailed_balance}
  W(u,\beta)=W(K-u,\beta) \exp(-\beta(K-2u))\ .
\end{equation}
The two best-known possibilities falling in this category are the
Metropolis rate,
\begin{equation}
W(u,\beta)= \min \left( 1,e^{-\beta(K-2u)} \right) \ , 
\end{equation}
and the Glauber rate,
\begin{equation}
W(u,\beta)= \frac{1}{2} 
\left(1 - \tanh\left(\beta\left(\frac{K}{2}-u\right)\right) \right) \ .
\end{equation}
We will keep, however, a general form for $W$ in our analytical
treatment, only assuming Eq. (\ref{eq:detailed_balance}) to be valid.
In the thermodynamic limit, this discrete process acquires a
continuous form by defining the time as $t=T/N$, and stepwise
differences of extensive observables translate into time derivatives
of the corresponding observable densities (extensive observables
divided by $N$).

The dynamics of the system, or more precisely of the probabilities
${\cal P}(\vec \sigma, t)$ that a microscopic configuration $\vec
\sigma$ is found at time $t$, can be completely described by the
master equation
\begin{equation}
  \label{eq:master}
  \frac{d}{dt} {\cal P}(\vec \sigma, t) = \frac 1N \sum_{i=1}^N \left[
  - W(u_i(\vec\sigma),\beta)\ {\cal P}(\vec \sigma,t) 
  + W(u_i(F_i \vec\sigma),\beta)\ {\cal P}(F_i \vec \sigma,t)
  \right]\ .
\end{equation}
It is, however, far too complicated to solve these $2^N$ coupled
equations directly. We are therefore going to present several
approximative characterizations of the dynamics of the model. They all
rely on the same idea: instead of following the evolution of the full
distribution ${\cal P}(\vec \sigma, t)$ of microscopic configurations,
we turn to a simplified description, in terms of a finite number of
macroscopic observables. The dynamic evolution of these cannot be
expected to be closed, as we have lost information with respect to the
microscopic description. It depends in general on a larger number of
macroscopic variables, i.e.~a hierarchical set of dynamic evolution
equations arises. Choosing carefully a closure hypothesis at any
level of this hierarchy, we are led to improvingly precise
predictions.

In the next three subsections the first three levels of this
hierarchy are presented, together with the corresponding closure assumptions.
The evaluation of their accuracy is deferred until Sec.~\ref{sec:mc}, where
we compare them with numerical simulations.

\subsection{The binomial approximation}
\label{subsec:bino}
The simplest implementation of this idea consists in keeping track of
the energy density $e(t)$ and of the magnetization per spin $m(t)$ only.

It is rather natural to include the energy in our set of observables.
Indeed, the system is evolving towards equilibrium with respect to the
Gibbs measure, in other words, at long times all the microscopic
configurations with equilibrium energy are equiprobable. Including the
magnetization is also necessary for a ferromagnetic system, where the
low temperature phase is characterized by a non zero value of the
magnetization. In a more general setting, the minimal set of
observables is given by the energy of the system and a complete set of
order parameters which allow for the exact description of the
equilibrium distribution.

At each time-step, the chosen spin has value $\sigma$ and $u$
unsatisfied edges around it with a certain probability
$p_{\sigma}(u;t)$.
If the flip is accepted, i.e. with probability $W(u,\beta)$, the total
energy changes by an amount of $K-2u$ (unsatisfied edges become
satisfied, and {\it vice versa}), and the variation of the total
magnetization is $-2\sigma$.  We thus obtain in the thermodynamic
limit:
\begin{eqnarray}
\frac{de}{dt} &=& \sum_{u=0}^K W(u,\beta) (K-2u) [p_{-}(u;t) + p_{+}(u;t)]\ , 
\label{eq:binomial_e} \\
\frac{dm}{dt} &=& 2 \sum_{u=0}^K W(u,\beta) [p_{-}(u;t) - p_{+}(u;t)] \ .
\label{eq:binomial_m}
\end{eqnarray}
Although exact, these equations are not of direct use because they involve
$p_{\sigma}(u;t)$, a dynamical quantity not present in our original
description via $\{e(t),m(t)\}$. We thus have to express approximately
$p_{\sigma}(u;t)$ in terms of $e$ and $m$ in order to close the set of
equations. The randomly selected variable has spin $\sigma$ with
probability $(1+\sigma m(t))/2$.  In the absence of further
informations, we can only assume that each of the $K$ incident edges
is unsatisfied with the same probability $\alpha_{\sigma}(t)$. This
yields the approximate binomial expression
\begin{equation}
p_{\sigma}(u;t) \simeq 
\frac{1+\sigma m(t)}{2} {K \choose u} \alpha_\sigma (t)^u 
(1 - \alpha_\sigma (t))^{K-u} \ .
\end{equation}
$\alpha_\pm$ can be determined by the following consistency condition.
All unsatisfied edges connect anti-parallel spins. The energy must
then be the same if expressed as the number of $+$ spins around $-$
spins, or the other way around.
\begin{eqnarray}
e(t)&=& \sum_{u=0}^K u \ p_+ (u;t) = \frac{1+m(t)}{2} K \alpha_+(t) 
\nonumber \\
&=& \sum_{u=0}^K u \ p_- (u;t) = \frac{1-m(t)}{2} K \alpha_-(t) \ .
\label{eq:consistency1}
\end{eqnarray}
We finally obtain
\begin{equation}
p_{\sigma}(u;t)=\frac{1+\sigma m(t)}{2} {K \choose u} \left(\frac{2e(t)}
{K(1+\sigma m(t))}\right)^u \left(1-\frac{2e(t)}{K(1+\sigma m(t))}\right)^{K-u}
\ .
\label{eq:binomial_p}
\end{equation}
This expression has the same form as the equilibrium equation
(\ref{eq:p_sigma_u2}), but with $e(t)$ and $m(t)$ being dynamical
variables, which can differ from their equilibrium values.

Equations (\ref{eq:binomial_e}), (\ref{eq:binomial_m}) and 
(\ref{eq:binomial_p}) can be condensed into
\begin{eqnarray}
\frac{de}{dt}&\equiv&F_e(e(t),m(t),\beta) \ ,\\
\frac{dm}{dt}&\equiv&F_m(e(t),m(t),\beta) \ .
\end{eqnarray}
A few expected properties of these equations can be checked
immediately:
\begin{itemize}
\item $F_m(e,m=0,\beta)=0$ for all $e$ and $\beta$. If the system is
  strictly unmagnetized, the dynamics does not break this symmetry in
  the thermodynamic limit. In the low temperature phase, there is of
  course an instability with respect to magnetization fluctuations
  (which are present in any finite system), as we shall show in Sec.
  \ref{sec:critical}.
\item $F_e(e_p,m=0,\beta)=0$ where $e_p=\frac{K}{2}(1+e^{\beta})^{-1}$
  is the paramagnetic energy density at inverse temperature $\beta$.
  To prove this, one can note that in that case
  \begin{equation}
    p_+(K-u) = e^{-\beta(K-2u)} p_-(u) \ .
    \label{eq:rel_static_p}
  \end{equation}
  Using the detailed balance condition on $W$ and the change of variables 
  $u \mapsto K-u$ in one of the sums defining $F_e$, one thus shows that the 
  paramagnetic state is a fixed point of the dynamics.
\item $F_e(e_f,m_f,\beta)=F_m(e_f,m_f,\beta)=0$ for $\beta>\beta_c$, where
  $m_f$ and $e_f$ are the magnetization and the energy density of the
  ferromagnetic phase.  Indeed, relation (\ref{eq:rel_static_p}) is
  still valid, as can be seen from (\ref{eq:p_sigma_u}), and the proof
  follows the same line as in the paramagnetic case: the ferromagnetic
  state, when present, is also a fixed point of the dynamics.
\end{itemize}

\subsection{Independent-neighbor approximation}
\label{sec:sites}

In the last section, we have described the dynamics of our
ferromagnetic model by mapping it to two coupled differential
equations for the energy and the magnetization. In order to approximately
close these equations we had to assume a specific binomial form for
the quantities $p_\sigma(u;t)$ which is exactly valid only in thermal
equilibrium, but not for intermediate times. It seems therefore
natural to extend the set of considered observables, and to look for
dynamical equations for $p_\sigma(u;t)$ itself.  Note that on regular
graphs, these quantities still form a finite set of observables. This
would not be the case for a graph with unbounded fluctuations of the
connectivities.

To formulate the dynamical equations, we have to take into account
different contributions to the variation of $p_\sigma(u;t)$ during a time-step:
\begin{itemize}
\item The first type of contribution is due to the flipped spin itself, i.e.
  $\sigma$ is flipped to $-\sigma$. All incident edges change from
  satisfied to unsatisfied and vice versa, i.e. we have
  $u\leftrightarrow K-u$.
\item The second type comes from the neighbors of the flipped
  spin. For these vertices, $\sigma$ remains obviously unchanged, but $u$ is
  increased or decreased by one depending on whether the connecting
  edge was satisfied or not before the flip.
\end{itemize}
Including the corresponding loss and gain terms, we find the following
exact equations,
\begin{eqnarray}
  \label{eq:site_dyn}
  \frac d{dt} p_\sigma (u;t) 
  &=& - W(u,\beta)\ p_\sigma(u;t) + W(K-u,\beta)\ p_{-\sigma}(K-u;t)
  \nonumber\\
  &&+\sum_{\tilde u}(K-\tilde u)\ W(\tilde u,\beta)\ p_\sigma(\tilde u;t)\   
  \left[- p(u | \sigma,\tilde u\to \sigma;t) + p(u-1 | \sigma,\tilde
    u\to \sigma;t) \right] \nonumber\\
  &&+\sum_{\tilde u} \tilde u\ W(\tilde u,\beta)\ p_{-\sigma}(\tilde u;t)\
  \left[- p(u | -\sigma,\tilde u\to \sigma;t) + p(u+1 |
    -\sigma,\tilde u\to \sigma;t) \right] \ ,
\end{eqnarray}
where $p(u | \tilde\sigma,\tilde u\to \sigma;t)$ is the conditional
probability that a vertex of spin $\sigma$ belongs exactly to $u$
unsatisfied edges, under the condition that this vertex is reached via
an edge coming from a vertex with spin $\tilde\sigma$ and with $\tilde
u$ unsatisfied bonds. This probability can be computed from the
joint probability $p_{\tilde\sigma\sigma}(\tilde u, u;t)$ of two adjacent
vertices which, in the equilibrium context, was already introduced at
the end of Sec.  \ref{sec:model}:
\begin{equation}
  \label{eq:joint_to_cond}
  p(u | \tilde\sigma,\tilde u\to \sigma;t) = \frac
  {p_{\tilde\sigma\sigma}(\tilde u, u;t)}
  {\sum_{\tilde u} p_{\tilde\sigma\sigma}(\tilde u, u;t)} \ . 
\end{equation}
As in the previous subsection, 
these exact equations do not close. 
The time evolution of the probabilities $p_\sigma(u;t)$ is given in terms of
the correlations between neighboring vertices,
$p_{\tilde\sigma\sigma}(\tilde u, u;t)$. We can, however, close the
equations at least approximately by considering neighbors as
independent \cite{BaHaWe},
\begin{eqnarray}
  \label{eq:independent}
  p(u |\sigma,\tilde u\to\sigma;t) &\simeq& \frac{(K-u)\
    p_{\sigma}(u;t)}{\langle K-u \rangle_\sigma}
  \nonumber\\ 
  p(u |-\sigma,\tilde u\to \sigma;t) &\simeq&  \frac{u \ p_\sigma
    (u;t)}{\langle u \rangle_\sigma}\ , 
\end{eqnarray}
i.e. the conditional probability does not depend on the properties of
the initial vertex, but only on the fact that we reach the new
vertex via a satisfied or an unsatisfied edge. In the last equation we
have used the short-hand notation $\langle\bullet\rangle_\sigma = \sum_u
\bullet\ p_\sigma(u;t)$. Note that this does not describe an average
since, for instance, $\langle 1\rangle_\sigma$ equals the fraction of spins
with value $\sigma$, and normalization holds only for
$\langle\bullet\rangle_++\langle\bullet\rangle_-$.

Under this approximation, Eq. (\ref{eq:site_dyn}) closes in $p_\sigma
(u;t)$ and becomes
\begin{eqnarray}
  \label{eq:site_dyn_closed}
  \frac d{dt} p_\sigma (u;t) 
  &=& - W(u,\beta)\ p_\sigma(u;t) + W(K-u,\beta)\ p_{-\sigma}(K-u;t)
  \nonumber\\
  &&+\frac{\langle(K-\tilde u)\ W(\tilde u,\beta)\rangle_\sigma}
  {\langle K-\tilde{u} \rangle_\sigma}  
  \left[- (K-u)\ p_\sigma(u;t) + (K-u+1) p_\sigma(u-1;t) \right]
  \nonumber\\ 
  &&+\frac{\langle \tilde u \ W(\tilde u,\beta)\rangle_{-\sigma}}
  {\langle \tilde u \rangle_\sigma}  
  \left[- u\ p_\sigma(u;t) + (u+1) \ p_\sigma(u+1;t) \right]  
\end{eqnarray}

An important observation is that Eq. (\ref{eq:independent}) becomes
exact for the equilibrium distributions (\ref{eq:p_sigma_u}) and
(\ref{eq:link_equ}). Therefore the true thermal equilibrium is a fixed
point of the closed dynamical equations. For intermediate times,
however, there will be deviations from Eq. (\ref{eq:site_dyn_closed}).
Since the description via $p_\sigma(u;t)$ is, however, more detailed
than the one of the binomial approximation, we expect deviations to be
less important. A thorough comparison with Monte-Carlo simulations
will be given in Sec. \ref{sec:mc}.

These equations are ordinary differential equations and can thus be
solved numerically by standard methods. One has, however, to be
careful with the choice of initial conditions. In fact, not every
normalized $p_\sigma(u)$ corresponds to microscopic configurations
$(\sigma_1,...,\sigma_N)$. A necessary and sufficient consistency
condition is given by the fact that each unsatisfied edge connects 
two antiparallel spins, see the discussion before Eq.
(\ref{eq:consistency1}). Therefore 
\begin{equation}
  \label{eq:consistency}
  \sum_u u\ p_+(u;t) =  \sum_u u\ p_-(u;t) 
\end{equation}
must hold for arbitrary time $t$. Restricting the allowed initial
conditions to all $p_\sigma(u;t)$ fulfilling this condition,
consistency is preserved by the dynamical evolution
(\ref{eq:site_dyn_closed}). This in fact
guarantees that the only stationary points of the dynamics are the
solutions of the equilibrium study of Sec. \ref{sec:model}. In the
high-temperature phase, the paramagnetic solution attracts the dynamics,
while the low-temperature phase allows for three stationary points.
The two ferromagnetic ones are stable, whereas the paramagnetic
solution is unstable with respect to any spin-flip asymmetry
($p_\sigma(u;t_0) \neq p_{-\sigma}(u;t_0)$) in the initial condition.

\subsection{Inclusion of neighbor correlations}

In order to further refine the description of the non-equilibrium
behavior of the model, we may ``iterate'' the step which led us from
the binomial to the independent-neighbor approximation. The exact
equations for the evolution of the single-site quantity
$p_\sigma(u;t)$ depend on the joint distribution $p_{\tilde \sigma
  \sigma}(\tilde u, u;t)$ for neighboring sites. Formulating an
equation for the time evolution of the latter quantity, which again
includes a finite set of functions, we find an equation including
three-spin correlations.  Approximately, these can be expressed in
terms of the two-spin distribution, and the dynamical equations close.
In the formulation of the equations we have to include the effects of
the flipped spin itself, and of its first and second neighbors. We
only give the resulting equations, where the time dependence is not
stated explicitly to lighten notations:
\begin{eqnarray}
  \label{eq:link-quantity}
  \frac d{dt} p_{\sigma\sigma} (u_1,u_2) &=& -W(u_1,\beta)\ 
  p_{\sigma\sigma}(u_1,u_2) + W(K-u_1,\beta)\ 
  p_{-\sigma\sigma}(K-u_1,u_2+1) \nonumber\\
  && + \sum_u W(u,\beta)\ \big[-p_{\sigma\sigma}(u,u_1)\ 
    (K-u_1-1)\ p(u_2|\sigma,u_1\to\sigma) \nonumber\\
  &&\ \ \ \ \ \ \ \ \ \ \ \ \ \ \ \ \ \ \ \ \
    +p_{\sigma\sigma}(u,u_1-1)\ (K-u_1)\
    p(u_2|\sigma,u_1-1\to\sigma) \nonumber\\
  &&\ \ \ \ \ \ \ \ \ \ \ \  \ \ \ \ \ \ \ \ \
    -p_{-\sigma\sigma}(u,u_1)\ (K-u_1)\
    p(u_2|\sigma,u_1\to\sigma) \nonumber\\
  &&\ \ \ \ \ \ \ \ \ \ \ \  \ \ \ \ \ \ \ \ \
    +p_{-\sigma\sigma}(u,u_1+1)\ (K-u_1-1)\ 
    p(u_2|\sigma,u_1+1\to\sigma)\ \big] \nonumber\\
  && + (u_1 \leftrightarrow u_2) \ , \nonumber\\
  \frac d{dt} p_{-\sigma\sigma} (u_1,u_2) &=&
  -\big[W(u_1,\beta)+W(u_2,\beta)\big] 
  p_{-\sigma\sigma}(u_1,u_2) \nonumber\\
  && + W(K-u_1,\beta)\ 
    p_{\sigma\sigma}(K-u_1,u_2-1) + W(K-u_2,\beta)\ 
    p_{-\sigma-\sigma}(u_1-1,K-u_2) \nonumber\\
  && + \sum_u W(u,\beta)\ \big[-p_{\sigma-\sigma}(u,u_1)\ 
    (u_1-1)\ p(u_2|-\sigma,u_1\to\sigma) \nonumber\\
  &&\ \ \ \ \ \ \ \ \ \ \ \ \ \ \ \ \ \ \ \ \
    +p_{\sigma-\sigma}(u,u_1+1)\ u_1\
    p(u_2|-\sigma,u_1+1\to\sigma) \nonumber\\
  &&\ \ \ \ \ \ \ \ \ \ \ \ \ \ \ \ \ \ \ \ \
    -p_{-\sigma-\sigma}(u,u_1)\ u_1\
    p(u_2|-\sigma,u_1\to\sigma) \nonumber\\
  &&\ \ \ \ \ \ \ \ \ \ \ \ \ \ \ \ \ \ \ \ \
    +p_{-\sigma-\sigma}(u,u_1-1)\ (u_1-1)\
    p(u_2|-\sigma,u_1-1\to\sigma) \nonumber\\
  &&\ \ \ \ \ \ \ \ \ \ \ \ \ \ \ \ \ \ \ \ \
    -p_{\sigma\sigma}(u,u_2)\ u_2\
    p(u_1|\sigma,u_2\to-\sigma) \nonumber\\
  &&\ \ \ \ \ \ \ \ \ \ \ \ \ \ \ \ \ \ \ \ \
    +p_{\sigma\sigma}(u,u_2-1)\ (u_2-1)\
    p(u_1|\sigma,u_2-1\to-\sigma) \nonumber\\
  &&\ \ \ \ \ \ \ \ \ \ \ \ \ \ \ \ \ \ \ \ \
    -p_{-\sigma\sigma}(u,u_2)\ (u_2-1)\
    p(u_1|\sigma,u_2\to-\sigma) \nonumber\\
  &&\ \ \ \ \ \ \ \ \ \ \ \ \ \ \ \ \ \ \ \ \ 
    +p_{-\sigma\sigma}(u,u_2+1)\ u_2\
    p(u_1|\sigma,u_2+1\to-\sigma)\ \big]\ ,
\end{eqnarray}
where $(u_1 \leftrightarrow u_2)$ means that the complete expression
on the right-hand side of the equation with $u_1$ and $u_2$
interchanged has to be added. In the following, these equations will
be denoted shortly as {\it link approximation}. As in the case of the
single-site quantities, they have to be solved numerically. The
$p_\sigma(u)$ can be recovered from the pair distribution in two
distinct ways:
\begin{equation}
  \label{eq:consistency_links}
  \sum_{u_2} p_{\sigma\sigma}(u_1,u_2) \propto (K-u_1)\ p_\sigma (u_1)    
\end{equation}
and 
\begin{equation}
  \label{eq:consistency_links2}
  \sum_{u_2} p_{\sigma-\sigma}(u_1,u_2) \propto u_1\ p_\sigma (u_1)\ .    
\end{equation}
These two procedures must yield the same value of $p_\sigma(u)$. 
This consistency condition ensures, in analogy to
Eq. (\ref{eq:consistency}) for the previous approximation scheme, that
the only stationary points are those given by the static approach. 

In principle, one could go on like this, i.e. equations for higher-order
correlations can be written down exactly. They will depend on even
higher correlations, and an infinite hierarchy of exact equations
arises. This hierarchy can be cut approximately at any arbitrary level
by factorizing higher correlations. Since the number of order
parameters used to describe the dynamics increases with each level, and
the observables of lower levels are contained via consistency
conditions, the description is expected to become more and more
precise. On the other hand, the equations become more and more
complex, as is already clear when one compares them for the 
three first levels of the
approximation hierarchy. Instead of continuing further in this way, we
compare now the results of the first three approximation levels with
Monte Carlo simulations. As we shall see, already these approximations
give an astonishing coincidence with numerical data.

\section{Comparison with numerics}
\label{sec:mc}

The approximation schemes presented in the last section do not possess
any intrinsic criterion to measure their quality. We therefore have
performed numerical simulations for large graphs in order
to check all three schemes.

To do so, we have first generated large random regular graphs
($N=3\cdot10^6$) of various connectivities, according to the algorithm
of \cite{Ne}. Then we have performed Monte Carlo simulations of the
ferromagnetic Ising model defined on this graph. In these simulations,
we have used as well Glauber as Metropolis dynamics, cf. Sec.
\ref{sec:defs}. In the following presentation we concentrate, however,
solely on the Metropolis case.  The results for Glauber dynamics
differ quantitatively, but the coincidence between numerical and
analytical results has the same quality. To suppress finite-size
fluctuations, which become important for large times close to the
critical point, we have averaged numerical data for up to 200
independent runs on independently generated graphs.

In general, we have initialized the system in the following way: all
spins are assigned randomly and independently a value, with a certain
bias in order to impose some initial magnetization on the system. Note
that this configuration is not an equilibrium configuration for any
arbitrary temperature. Then we have measured the relaxational dynamics
from this non-equilibrium configuration to thermal equilibrium, and
recorded in particular the time evolution of the energy density $e(t)$
and of the global magnetization $m(t)$.

\begin{figure}
  \begin{center}
    \includegraphics[angle=0,width=8cm]{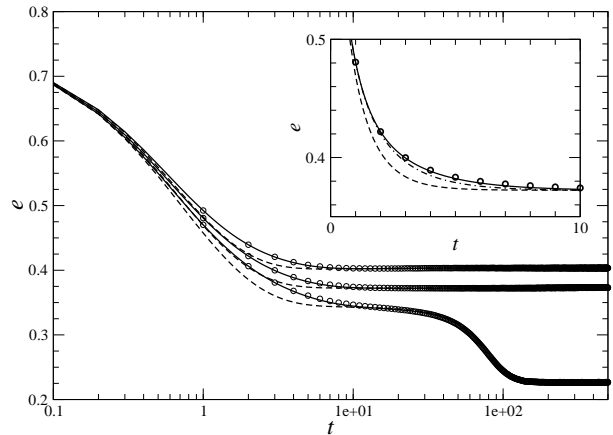}\\[0.3cm]
    \caption{Energy density as a function of time. The three approximations
      (dashed line = binomial, dash-dotted = independent neighbor,
      full = link approximation) are compared to numerical simulations
      (symbols, $N=3\cdot10^6,\ K=3$, averaged over 200 runs, error
      bars are smaller than the symbol size) for inverse temperatures
      $\beta=1.0, \ln 3, 1.2$ (top to bottom), with critical value
      $\beta_c=\ln 3$.  In the initial condition all spins were drawn
      independently, with average magnetization 0.1. In the main
      plot, the independent neighbor approximation is not shown
      because, on the scale of the figure, it is very close to the
      link approximation. The shoulder in the curve for $\beta=1.2$ is
      located close to the paramagnetic energy value, the relaxation
      to the ferromagnetic state appears on a longer time scale. The
      inset enlarges the region of largest deviation between the
      different approximations for $\beta_c$. Obviously, the more
      involved schemes lead to better approximations, the difference
      between the link approximation and the numerical data is hardly
      visible.}
    \label{fig:energy}
  \end{center}
\end{figure}

\begin{figure}
  \begin{center}
    \includegraphics[angle=0,width=8cm]{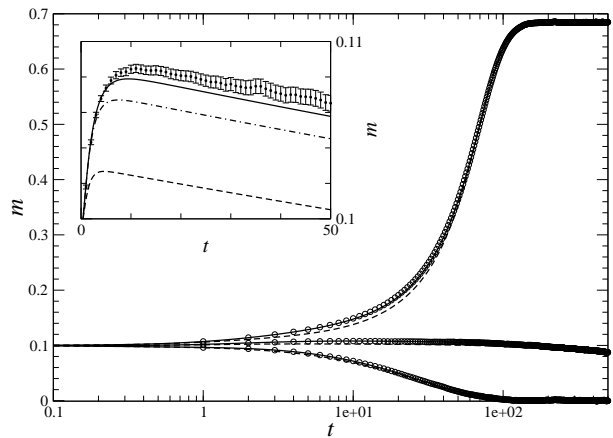}\\[0.3cm]
    \caption{Magnetization as a function of time. The parameters and
      symbols are the same
      as in Fig. \ref{fig:energy}. Again, the coincidence of
      the link approximation with the numerical data is excellent for
      the full time interval.
    }
    \label{fig:magnetization}
  \end{center}
\end{figure}

As a result, as represented in Figs. \ref{fig:energy} and
\ref{fig:magnetization}, we have found the following behavior: for
very short and very long time, all three approximation schemes are in
extremely good coincidence with numerical data. This is even true
close to the critical temperature, all three schemes reproduce with 
high precision the critical slowing down, i.e. the diverging longest
time scale in the system. Exactly at the critical point, all three
approximation schemes show the correct algebraic relaxation towards
equilibrium. So even the simplest approximation, which can be analyzed
in large detail, see the next section, allows for a precise estimate
of the relaxational dynamics close to equilibrium.

As shown in the insets of Figs. \ref{fig:energy} and
\ref{fig:magnetization}, considerable deviations between numerical
data and analytical predictions appear only for intermediate times.
As to be expected, these deviations become smaller for more detailed
approximations. In the link-approximation, these deviations are
hardly visible for all times, even at the critical point.

In addition, we have observed an increasing precision of all three
approximations for growing $K$. We expect therefore, that the binomial
approximation becomes exact in the large-$K$ limit \cite{SeMo}.

\section{Critical behavior}
\label{sec:critical}

In this section we investigate in more detail the predictions
of our approximations in the critical region of temperatures
separating the paramagnetic and the ferromagnetic phase. For the sake
of simplicity, we concentrate on the first two levels of
approximations, for which more detailed computations can be done
explicitly.

As we have seen before, all the presented dynamical approximations have
fixed points corresponding to paramagnetic and ferromagnetic
equilibrium. One has to study now the nature of the dynamical flows in
the space of projected order parameters. At high temperatures, the
only fixed point is the paramagnetic one, and it is obvious on
physical grounds that it will be stable. In the low temperature phase,
this fixed point will become unstable and the two ferromagnetic ones
will attract the dynamics.

Quite generally, if the projected evolution is described by an
$n$-component vector $\vec{q}(t)$ of observables, the system's
evolution is given by $n$ equations $\dot{q}_i=F_i(\vec{q})$ . Fixed
points correspond to $F_i(\vec{q}_0)=0 \ \forall i$. They are locally
stable if and only if all eigenvalues of the
$n \times n$ matrix $\cal M$ are negative, where $\cal M$ is defined by
its entries ${\cal M}_{ij}=(\partial F_i)/(\partial q_j)$, evaluated
at the fixed point under consideration. In this case, the asymptotic
relaxation of a fluctuation towards equilibrium is given by the
largest eigenvalue $\lambda_{max}$ (smallest in absolute value), and
the longest relaxation-time scale of the model reads
\begin{equation}
  \label{eq:relaxtimescal}
  \tau = - (\lambda_{max})^{-1}\ .
\end{equation}
Approaching the critical point, this eigenvalue tends to zero. The
system slows down until the relaxation time eventually diverges at
$\beta_c$. Right at the critical point, fluctuations decay only
algebraically with time.

\subsection{Relaxation time and critical slowing down in the binomial
  approximation} 

In the simplest case of the binomial approximation, the space of
parameters is only two-dimensional, the state of the system being
characterized by its magnetization and energy. Following the
notations of Sec. \ref{subsec:bino}, we define the matrix $\cal M$ by
\begin{equation}
{\cal M} = \left( \begin{array}{c c} {\cal M}_{ee} & {\cal M}_{em} \\
 {\cal M}_{me} & {\cal M}_{mm} \end{array} \right)
=\left( \begin{array}{c c} \frac{\partial F_e}{\partial e} &
\frac{\partial F_e}{\partial m} \\
\frac{\partial F_m}{\partial e} & \frac{\partial F_m}{\partial m}
\end{array} \right) \ .
\end{equation}
This matrix takes a particularly simple form when computed at the paramagnetic
fixed point. Indeed, the symmetry of the system under magnetization reversal
causes the non-diagonal elements to vanish, and one can directly read the 
eigenvalues of $\cal M$ in the diagonal entries,
\begin{eqnarray}
{\cal M}_{ee}&=& - \frac{1}{K} \frac{1+e^\beta}{(1+e^{-\beta})^{K-1}} 
\sum_{u=0}^K {K \choose u} W(u,\beta) e^{-\beta u} (K-2u)^2 
\ , \\
{\cal M}_{mm}&=& - \frac{2-K+K e^{-\beta}}{(1+e^{-\beta})^K}
\sum_{u=0}^K {K \choose u} W(u,\beta) e^{-\beta u}
\ .
\end{eqnarray}
 
It is obvious from these expressions that
${\cal M}_{ee}$ is negative at all temperatures, 
and that ${\cal M}_{mm}$ changes its sign at the critical
temperature $\beta_c=\ln(K/(K-2))$.
At low temperature, the paramagnetic fixed point becomes unstable 
against fluctuations of the magnetization, as expected on physical
grounds.

From the above expression of ${\cal M}_{mm}$ one obtains the divergence
of the relaxation time of the system when approaching the critical
temperature from above,
\begin{equation}
\tau_p \underset{\beta \to \beta_c^-}{\sim} (\beta_c-\beta)^{-1} 
\left[ \frac{K-2}{(1+e^{-\beta_c})^K}
\sum_{u=0}^K {K \choose u} W(u,\beta_c) e^{-\beta_c u} \right]^{-1} \ .
\label{divergence_para}
\end{equation}

In the low temperature phase,
at the ferromagnetic fixed point (we consider only the one with positive 
magnetization for simplicity), the four elements are different from zero.
Both eigenvalues of $\cal M$ are negative, implying the stability of 
the ferromagnetic 
phase. As their expressions at general temperature is not very illuminating, 
we concentrate on the $\beta \to \beta_c^+$ limit, for which one finds
the following scaling of the matrix elements:
\begin{equation}
{\cal M}_{ee} \to a <0 \; \; , \; \; 
{\cal M}_{em} \sim b \sqrt{\beta - \beta_c} \; \; , \; \; 
{\cal M}_{me} \sim c \sqrt{\beta - \beta_c} \; \; , \; \;
{\cal M}_{mm} \sim d (\beta - \beta_c) \ ,
\end{equation}
where $a$, $b$, $c$ and $d$ can be explicitly computed in terms of 
$W(u,\beta_c)$.
This implies that the relevant eigenvalue vanishes as 
\begin{equation}
\lambda \sim \left( d - \frac{bc}{a} \right) (\beta - \beta_c) \ .
\end{equation}
After some algebra to simplify the expression, we obtain finally the
divergence of the relaxation time from the ferromagnetic side of the transition
as
\begin{equation}
\tau_f \underset{\beta \to \beta_c^+}{\sim} (\beta-\beta_c)^{-1} 
\left[ \frac{2(K-2)}{(1+e^{-\beta_c})^K}
\sum_{u=0}^K {K \choose u} W(u,\beta_c) e^{-\beta_c u} \right]^{-1} \ .
\label{divergence_ferro}
\end{equation}
The two expressions (\ref{divergence_para}) and (\ref{divergence_ferro}) 
show an universal amplitude ratio of $1/2$.

\subsection{Relaxation time and critical slowing down in the 
  independent-\\neighbor approximation} 

The analysis becomes more involved in the independent-neighbor
approximation. As said above, we have to compute the $2(K+1)\times
2(K+1)$ matrix
\begin{equation}
  \label{eq:Mina}
  {\cal M} = \left( \frac{\partial \dot p_{\sigma_1}(u_1;t)}
  {\partial p_{\sigma_2}(u_2;t)} \right)
\end{equation}
starting from equation (\ref{eq:site_dyn_closed}), evaluated at the
equilibrium distribution $p_{\sigma_2}(u_2)$ given in
(\ref{eq:p_sigma_u}). The matrix elements are thus given by
\begin{eqnarray}
  \label{eq:Mentries}
  {\cal M}_{\sigma,\sigma} (u_1,u_2) &=& - W(u_1,\beta)\
  \delta_{u_1,u_2} \nonumber\\
  && + \frac{ \langle (K-u) W(u,\beta) \rangle_\sigma} 
  { \langle K-u \rangle_\sigma} \left[ - (K-u_1)\ \delta_{u_1,u_2} +
    (K-u_1+1)\ \delta_{u_1-1,u_2}  \right]\nonumber\\
  && + \left[ -(K-u_1) p_\sigma(u_1) + (K-u_1+1) p_\sigma(u_1-1)
  \right] \nonumber\\
  && \ \ \ \times \left[ \frac{(K-u_2) W(u_2,\beta)}{\langle
      K-u\rangle_\sigma} 
  - \frac{\langle (K-u) W(u,\beta) \rangle_\sigma \ (K-u_2)}
  {\langle K-u\rangle_\sigma^2} \right]  \nonumber\\
  && + \frac{ \langle u W(u,\beta) \rangle_{-\sigma}} 
  { \langle u \rangle_\sigma} \left[ - u_1\ \delta_{u_1,u_2} +
    (u_1+1)\ \delta_{u_1+1,u_2}  \right]\nonumber\\
  && + \left[ -u_1 p_\sigma(u_1) + (u_1+1) p_\sigma(u_1+1)
  \right] \frac{\langle u W(u,\beta) \rangle_{-\sigma} \ u_2}
  {\langle u\rangle_\sigma^2}\ , \nonumber\\
  {\cal M}_{\sigma,-\sigma} (u_1,u_2) &=& W(K-u_1,\beta)\
  \delta_{K-u_1,u_2} \nonumber\\
  && + \frac{ u_2 W(u_2,\beta) }{ \langle u \rangle_\sigma} 
  \left[ -u_1 p_\sigma(u_1) + (u_1+1) p_\sigma(u_1+1) \right]
  \ .\nonumber\\
\end{eqnarray}
The relaxation times equal minus the inverse eigenvalues of this
matrix. One has, however, to be careful since not all eigenvectors
correspond to physically allowed deviations of $p_\sigma(u;t)$ from
its equilibrium value. As already discussed in Sec.~\ref{sec:sites},
the values of $p_\sigma(u)$ are restricted by normalization and by the
consistency condition (\ref{eq:consistency}). The corresponding
eigenvectors, one being proportional to $p_\sigma(u)$ itself, the
other one to the deviation $\partial p_\sigma(u) / \partial h$
according to a deviation of the effective field $h$ away from its
self-consistent cavity value, correspond to zero eigenvalues of ${\cal
  M}$, and have to be excluded.

The most efficient way to achieve this is to explicitly require
normalization and consistency by, e.g., expressing $p_\pm (K;t)$
through the other values $p_\pm (u;t)$ with $u<K$. From normalization
and consistency we immediately find
\begin{equation}
  \label{eq:pk}
  p_\sigma (K;t) = \frac 12 \left[ 1- \sum_{u=0}^{K-1} \left(
      \frac{K+u}K p_\sigma(u;t) +  \frac{K-u}K p_{-\sigma}(u;t)
    \right) \right]\ .
\end{equation}
The matrix ${\cal M}$ becomes thereby reduced to a $2K$-dimensional
matrix $\tilde{\cal M}$ with entries $(u_{1,2}=0,...,K-1)$
\begin{eqnarray}
  \label{eq:tM}
  \tilde{\cal M}_{\sigma_1,\sigma_2}(u_1,u_2)
  &=& \frac{\partial \dot p_{\sigma_1}(u_1;t)}{\partial
    p_{\sigma_2}(u_2;t)} + \sum_{\sigma=\pm 1} \frac{\partial \dot
    p_{\sigma_1}(u_1;t)}{\partial p_{\sigma}(K;t)}\ \frac
  {\partial p_{\sigma}(K;t)}{\partial p_{\sigma_2}(u_2;t)}
 \nonumber\\ 
  &=& {\cal M}_{\sigma_1,\sigma_2}(u_1,u_2) - \sum_{\sigma=\pm 1}
  \frac{K-u\sigma_1\sigma}{2K}\  {\cal M}_{\sigma_1,\sigma}(u_1,K)\ .
\end{eqnarray}
Besides the excluded unphysical eigenvectors, this matrix has the same
eigenvalues and eigenvectors as ${\cal M}$.

Unfortunately, we were not able to find a general expression for the
smallest eigenvalue of $\tilde{\cal M}$ for arbitrary temperature, and
thus of the longest relaxation time of the system. For small values of
$K$, the latter can, however, easily be evaluated using a standard
computer-algebra system. We find, in complete accordance with the
binomial approximation, that the longest relaxation time scale
diverges like $\tau_p = A_K (\beta_c-\beta)^{-1}$ if the critical
point is reached from the paramagnetic side, and like $\tau_f = \frac
12 A_K (\beta-\beta_c)^{-1}$ coming from lower temperature. We thus
find the same critical exponent and the same universal amplitude ratio
$1/2$. Only the prefactors are slightly modified, with a difference of
about 1-2\% only between the estimates of the binomial and of the
independent-neighbor approximations. The values for Metropolis
dynamics are recorded in the table below.

\begin{center}
\
\begin{tabular}{|c||c|c|}\hline
\label{tab:entropie}
$K$ & $A_K$ (binomial) & $A_K$ (independent neighbor)  \\ 
\hline \hline
3 & $\frac{16}5 = 3.2$ & $ \frac{1882}{557} \simeq 3.273043$ \\  
\hline
4 & $\frac{27}{28}\simeq 0.964286$ & $\frac{1146337}{1162000}\simeq
0.986521$  \\   
\hline
5 & 0.605604  & 0.609906  \\
\hline
6 & 0.393775 & 0.398242  \\  
\hline 
7 & 0.309897 & 0.311649  \\
\hline
\end{tabular}
\end{center}

\begin{figure}
  \begin{center}
    \includegraphics[angle=0,width=8cm]{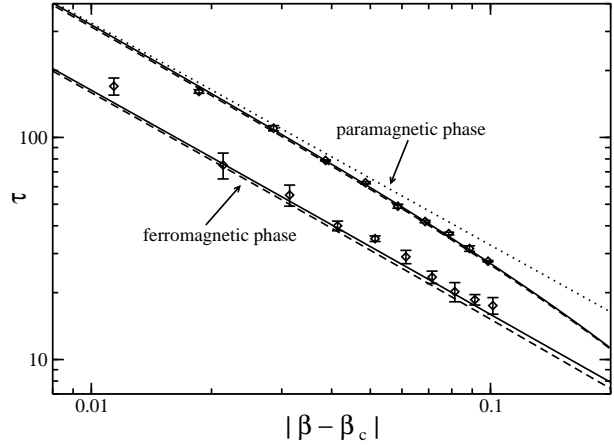}\\[0.3cm]
    \caption{Metropolis relaxation time for $K=3$ in
      the paramagnetic (top) and the ferromagnetic (bottom) phase. We
      have plotted the results of the binomial (dashed line) and the
      independent-neighbor (full line) approximations as extracted
      from the exact eigenvalues of the matrices ${\cal M}$. The
      dotted line gives the asymptotic algebraic divergence in the
      paramagnetic phase. The symbols are extracted from numerical
      simulations ($N=3\cdot 10^6$, 20 samples).  }
    \label{fig:relax_time}
  \end{center}
\end{figure}

In Fig.~\ref{fig:relax_time}, the results of both approximations are
compared to numerical data. The results of the independent neighbor
approximation are slightly higher than the binomial ones. In perfect
agreement with this observation, also the numerical data are found to
systematically deviate towards slightly higher relaxation times. Note,
however, that the extraction of the relaxation time in the
ferromagnetic phase is slightly subtle: the time interval between the
pre-asymptotic and the fluctuation-dominated dynamics is rather short
even for large systems ($N=3\cdot10^6$, averaged over 20 samples).

\subsection{Algebraic relaxation at criticality}

At the critical point $\beta_c$, the longest time scale
diverges, and the observables of the system decay only algebraically
towards their equilibrium values. Within the binomial approximation,
the dynamical exponents for energy and magnetization decay as well as
their prefactors can be determined analytically.

The equilibrium at the critical point is given by a non-magnetized
state of energy density $e_c=\frac{K(K-2)}{4(K-1)}$. If the initial
condition has vanishing magnetization, the energy relaxes
exponentially towards $e_c$. However, if the evolution starts from an
initial condition of arbitrarily small, but non-zero magnetization,
the dynamical evolution shows a power-law dependence in time.

Let us denote the excess-energy density by $\hat{e}=e-e_c$, and
expand the evolution equations around their fixed point $(e_c,m=0)$.  
Exploiting the spin-flip symmetry $m \leftrightarrow -m$ and
the fact that $\partial_m F_m$ vanishes at the critical temperature, we
obtain for the lowest orders
\begin{eqnarray}
\frac{d\hat{e}}{dt} &=& C^{(e)}_{10} \hat{e} + C^{(e)}_{02} m^2 +
C^{(e)}_{20} \hat{e}^2 + C^{(e)}_{12} \hat{e}m^2 + \dots \ , 
\\
\frac{dm}{dt} &=& C^{(m)}_{11} \hat{e} m + C^{(m)}_{03} m^3 +
C^{(m)}_{21} \hat{e}^2 m + C^{(m)}_{13} \hat{e} m^3 + C^{(m)}_{05} m^5
+ \dots \ , 
\end{eqnarray}
where the coefficients are given by
\begin{equation}
  \label{eq:pref}
  C^{(m/e)}_{ij} = \frac 1{i!\ j!} \left. \frac{\partial ^ {i+j}
    F_{m/e}(e,m,\beta_c)} 
  {\partial^i e\ \partial^j m} \right|_{e=e_c,\, m=0}  \  .
\end{equation}
To extract the leading algebraic long-time behavior
\begin{equation}
m(t) \sim m_0\ t^{-z_m} \qquad , \qquad \hat{e} \sim \hat{e}_0\
t^{-z_e} \ , 
\end{equation}
we have to compare the dominant terms on both sides of the equations.
Seen that the lhs of the first equation is of ${\cal O}(t^{-z_e-1})$,
the asymptotically larger ${\cal O}(t^{-z_e})$-term on the rhs has to
be compensated by the second contribution with ${\cal O}(t^{-2z_m})$.
This results in $z_e = 2 z_m$. The lhs of the second equation is of
${\cal O}(t^{-z_m-1})$, whereas the two dominant terms on its rhs are
of ${\cal O}(t^{-z_e-z_m})= {\cal O}(t^{-3z_m})$. Comparing the order of
these terms we consequently find
\begin{equation}
z_m = \frac 12 \qquad , \qquad z_e = 1 \ .
\end{equation}
These exponents stand in perfect agreement with numerical simulations.
Considering in addition the coefficients of the discussed
contributions, we are led to
\begin{eqnarray}
0 &=& \hat{e}_0 \frac{\partial F_e}{\partial e} + \frac{1}{2} m_0^2 
\frac{\partial^2 F_e}{\partial m^2} \ , \\
-\frac{1}{2} m_0 &=& 
 \hat{e}_0 m_0 \frac{\partial^2 F_m}{\partial e \partial m}
+ \frac{1}{6} m_0^3 \frac{\partial^3 F_m}{\partial m^3} \ ,
\end{eqnarray}
where the derivatives have to be evaluated at the fixed point
$(e_c,m=0)$.  After some algebra, we obtain the prefactors of
magnetization and excess energy:
\begin{eqnarray}
m_0^2 &=& \frac{3K}{2(K-2)} \left( \frac{2(K-1)}{K} \right)^{K-1} 
\left[\sum_{u=0}^K {K \choose u} W(u,\beta_c) e^{-\beta_c u} 
\right]^{-1} \ , \\
\hat{e}_0 &=& - \frac{K-2}{4} m_0^2 \ . 
\end{eqnarray}
Note that there are two solutions $m_0=\pm \sqrt{m_0^2}$, depending of
the sign of the magnetization in the initial condition. The quality of the 
agreement between these predictions and numerical simulations (not shown) is 
comparable to the one of previous subsection.

\section{Two-time correlations}
\label{sec:twotime}

We briefly sketch in this section an extension of the previously
introduced approximations to the study of two-time quantities, more
precisely of the global auto-correlation function of the spins,
\begin{equation}
C(t_2,t_1)=\frac{1}{N} \sum_{i=1}^N \sigma_i(t_2) \sigma_i(t_1) \ .
\end{equation}
These will be, in the thermodynamic limit, sharply peaked around its
average value (with respect to the possible histories of the
microscopic dynamics).  We suppose in the following $t_2 \ge t_1$ to
simplify the notations.  To compute this function, we shall consider
$t_2$ as the evolution time and project the dynamics onto a global
observable which retains a trace of the microscopic configuration at
the earlier time $t_1$. More precisely, let us call $q_{\sigma_1
  \sigma_2}(uu,us,su;t_1,t_2)$ the fractions of sites whose spin
equals $\sigma_1$ at time $t_1$ and $\sigma_2$ at time $t_2$, and
which have around them
\begin{itemize}
\item $uu$ edges which are unsatisfied at both times, 
\item $us$ which are unsatisfied at time $t_1$ and satisfied at time
  $t_2$,
\item $su$ which are satisfied at time $t_1$ and unsatisfied at time
  $t_2$,
\item and $ss \equiv K-uu-us-su$ which are satisfied at both times. 
\end{itemize}
We introduce the notation
\renewcommand{\arraystretch}{.75}
\begin{equation}
\langle \bullet \rangle_{\sigma_1 \sigma_2} =
\underset{\begin{array}{c} uu,us,su \\
  uu+us+su \le K \end{array}}{\sum} \bullet \; q_{\sigma_1
\sigma_2}(uu,us,su;t_1,t_2) \ ,  
\end{equation}
which is again not a normalized average for a given value of
$\sigma_1$ and $\sigma_2$, but only when the sum is taken over all
indices, $\langle 1 \rangle_{++} + \langle 1 \rangle_{+-} +\langle 1
\rangle_{-+} +\langle 1 \rangle_{--} =1$.  The two-time correlation of
the spins can be obtained from $q$, and reads in this short-hand
notation:
\begin{equation}
C(t_2,t_1) = \langle 1 \rangle_{++} + \langle 1 \rangle_{--} 
- \langle 1 \rangle_{+-} - \langle 1 \rangle_{-+} \ .
\label{twotime-correl}
\end{equation}

As in the previous cases, one has to impose consistency equations on
$q$, to ensure that the number of unsatisfied edges around up or down
spins is the same. Enforcing this condition at $t_1$ and $t_2$ yields
respectively:
\begin{equation}
\langle uu + us \rangle_{++} + \langle uu + us \rangle_{+-} = 
\langle uu + us \rangle_{-+} + \langle uu + us \rangle_{--} \ .
\end{equation}
\begin{equation}
\langle uu + su \rangle_{++} + \langle uu + su \rangle_{-+} = 
\langle uu + su \rangle_{+-} + \langle uu + su \rangle_{--} \ ,
\end{equation}

Boundary conditions also constrain the value of $q$ when $t_1=t_2$.
Obviously $q_{+-}=q_{-+}=0$ at equal times, and 
\begin{equation}
q_{\sigma \sigma}(uu,us,su;t,t)=\delta_{us,0} \; \delta_{su,0} \; 
p_\sigma(uu;t) \ . 
\label{twotime-boundary}
\end{equation}

An evolution equation with respect to $t_2$ for $q_{\sigma_1
  \sigma_2}(uu,us,su;t_1,t_2)$ can be closed using factorization
approximations similar to those used in Sec.\ref{sec:sites}, here we
only state the result:
\begin{eqnarray}
&&\frac{d}{dt_2} q_{\sigma_1 \sigma_2}(uu,us,su;t_1,t_2)= \nonumber \\ 
&-& q_{\sigma_1 \sigma_2}(uu,us,su) W(uu+su)
+ q_{\sigma_1, -\sigma_2}(us,uu,ss) W(ss+us) \nonumber \\ 
&+&\frac{\langle uu W(uu+su) \rangle_{-\sigma_1-\sigma_2}}{\langle uu
  \rangle_{\sigma_1 \sigma_2}} [-uu \; q_{\sigma_1
    \sigma_2}(uu,us,su) + (uu+1) q_{\sigma_1 \sigma_2}(uu+1,us-1,su)
] \nonumber \\ 
&+&\frac{\langle us W(uu+su) \rangle_{-\sigma_1 \sigma_2}}{\langle us
  \rangle_{\sigma_1 \sigma_2}} [ -us \; q_{\sigma_1
    \sigma_2}(uu,us,su) + (us+1) q_{\sigma_1 \sigma_2}(uu-1,us+1,su)
] \nonumber \\ 
&+&\frac{\langle su W(uu+su) \rangle_{\sigma_1 -\sigma_2}}{\langle su
  \rangle_{\sigma_1 \sigma_2}} [ -su \; q_{\sigma_1
    \sigma_2}(uu,us,su) + (su+1) q_{\sigma_1 \sigma_2}(uu,us,su+1)
] \nonumber \\ 
&+&\frac{\langle ss W(uu+su) \rangle_{\sigma_1 \sigma_2}}{\langle ss
  \rangle_{\sigma_1 \sigma_2}} [ -ss \; q_{\sigma_1
    \sigma_2}(uu,us,su) + (ss+1) q_{\sigma_1 \sigma_2}(uu,us,su-1)
] \ ,
\label{twotime-evolution}
\end{eqnarray}
where we suppressed the explicit time dependence of $q$ in the
right-hand side to lighten notations.

\begin{figure}
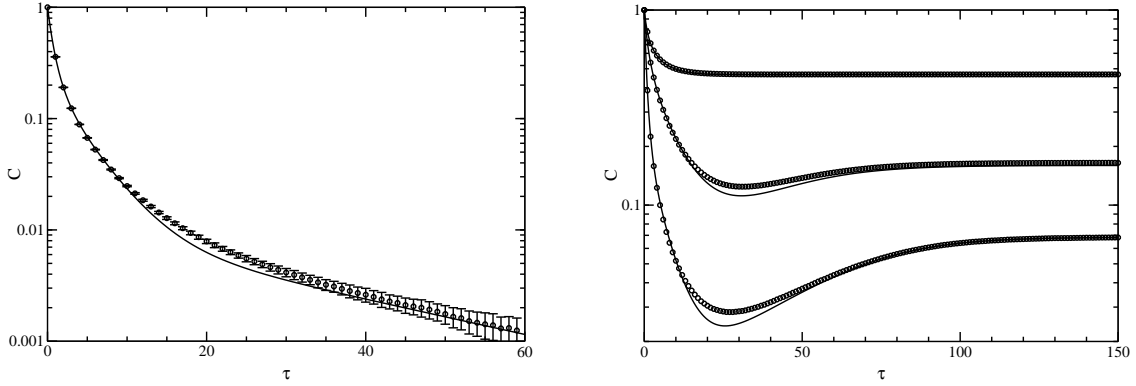

  \begin{center}
    \epsfig{file=correl_b1.eps,width=7cm} \hspace{7mm}
    \epsfig{file=correl_b12.eps,width=7cm}
  \end{center}
  \caption{Two-time correlations $C(t_1+\tau,t_1)$ as a function of $\tau$,
    for $K=3$. Solid lines: numerical integration of the differential
    equations, see the text for details. Symbols: Monte Carlo
    simulations.  Left: $\beta=1$, $t_1=0$. Numerical simulations
    averaged on 200 samples of size $N=10^7$. Right: $\beta=1.2$,
    $t_1=0$ (bottom), $30$ (middle), $150$ (top).  Numerical
    simulations averaged on 200 samples of size $N=3 \ 10^6$, error
    bars are smaller than the symbol size.}
  \label{fig:twotime}
\end{figure}

One can easily exploit these equations to compute the two-time
correlation $C(t_2,t_1)$ for a given initial configuration at $t=0$.
The first task is to determine $p_\sigma(u;t_1)$ by a numerical
integration of the equations (\ref{eq:site_dyn_closed}) between $t=0$
and $t=t_1$.  Using then the boundary condition
(\ref{twotime-boundary}) and the evolution equations
(\ref{twotime-evolution}), one obtains $q$ and finally $C$ through
(\ref{twotime-correl}). We confront in Fig. \ref{fig:twotime} the
results of such a procedure with Monte Carlo numerical experiments,
which are in satisfying agreement with each other.

\section{Connection with dynamical replica theory}
\label{sec:drt}

\subsection{Basic assumptions of dynamical replica theory}

In this section, we are going to show the equivalence of our
approximations to the dynamical replica theory (DRT)
\cite{CoSh,LaCoSh} under the assumption of replica symmetry. To do so,
we will derive the closure equations we used in Sec.~\ref{sec:dyn} 
starting from the assumptions of the dynamical replica approach.

Similar to our approach, DRT aims at an approximate description of the
non-equilibrium dynamics of disordered systems by means of projecting
the dynamics onto an array $\vec q$ of $n$ observables, and by
approximately closing the corresponding equations.  The method becomes
exact if the following two properties are fulfilled:
\begin{itemize}
\item[{\it (i)}] The observables in $\vec q$ have self-averaging 
  properties along
  the dynamics, i.e. they follow their average trajectories with
  probability one in the thermodynamic limit. 
\item[{\it (ii)}] At each time, all microscopic configurations $\vec
  \sigma = (\sigma_1,....,\sigma_N)$ having the same values of all
  observables in $\vec q$ are equi-probable.
\end{itemize}
The first assumption is not expected to pose serious problems if,
e.g., densities or fractions of vertices with certain characteristics
are considered.  The second assumption, on the other hand, is much
more crucial: it is obviously true in thermal equilibrium if the
energy density is included in $\vec q$. Indeed the
probability distribution of microscopic configurations is then given by the
Boltzmann-Gibbs distribution, i.e. it depends on the actual
configuration only via one observable, the energy. There is, however,
no reason why this should hold far from equilibrium. In fact, as
discussed before, we have observed that the inclusion of more and more
sophisticated observables gives a better and better description of the
dynamics of the system.

At variance with the original DRT, where replicas are introduced to
average over the disorder of the Sherrington-Kirkpatrick model
\cite{ShKi}, we are going to use the cavity method \cite{MePa}. The
latter is more efficient to average over the quenched disorder of
diluted systems, in our case represented by the random regular graph
on which the Ising model is defined. Even if this average is almost
trivial in the case of a random regular graph, where the disorder
enters only via large loops, i.e. via self-consistency conditions,
this method can be easily extended to fluctuating connectivities or
random interaction strengths.

\subsection{The binomial approximation}

Let us start the discussion with the binomial approximation, i.e. with
the case where the dynamics is completely approximated by the
behavior of the energy and the magnetization densities. The two
assumptions of DRT stated above result in the following approximation,
\begin{equation}
  \label{eq:bin_drt}
  {\cal P}(\vec \sigma, t) = \frac
  {\delta\left(m(t) - \frac 1N \sum_i \sigma_i \right)
   \ \delta\left(e(t) - \frac 1N H(\vec \sigma) \right)}
  {\sum_{\vec\sigma}\ \delta\left(m(t) - \frac 1N \sum_i \sigma_i
    \right) 
   \ \delta\left(e(t) - \frac 1N H(\vec \sigma) \right)}\ .
\end{equation}
Instead of working directly in this generalized micro-canonical
framework, we use a generalized canonical approach introducing
conjugate parameters for the energy, i.e. a formal inverse temperature
$\beta(e,m)$, and for the magnetization, i.e. a formal external field
$\gamma(e,m)$. We thus replace Eq. (\ref{eq:bin_drt}) by
\begin{equation}
  \label{eq:bin_drt_canonical}
  {\cal P}(\vec \sigma, t) = \frac 1{Z(\beta(e,m),\gamma(e,m))}
  \exp\left\{ -\beta(e,m) H(\vec \sigma)-\gamma(e,m) \sum_i \sigma_i  \right\}\
  .  
\end{equation}
Note that the formal temperature $\beta(e,m)$ is different from the
physical temperature $\beta$, as long as the system is not
equilibrated. Both conjugate parameters have to be adjusted such that
the average values of the energy and magnetization density assume the
desired values $e(t)$ and $m(t)$. In the thermodynamic limit, the
measure becomes sharply concentrated around these values.

Using these weights for the microscopic configurations, we have to prove 
that Eq. (\ref{eq:binomial_p})
holds, i.e. that the two assumptions of DRT lead to the same closure
of our approximate dynamical equations. This can be easily obtained
using the Bethe-Peierls approach sketched in Sec. \ref{sec:model}, the
only modification is due to the additional external field. The
analysis follows, however, exactly the same steps, and it leads in
particular to the desired binomial closure assumption.

\subsection{The independent-neighbor approximation}

The case of the independent neighbor approximation is only slightly
more involved. We have to show that the assumption that all
configurations with the same $p_\sigma(u)$ are equiprobable leads to
the desired factorization of the joint distribution of neighbors, and
thus to the dynamical equations of Sec. \ref{sec:sites}. More
precisely, we demonstrate that, under the above-stated assumption, the
conditional probability of finding a vertex with $u_2$ unsatisfied
edges, following a $\sigma\to\pm\sigma$ edge from a vertex with $u_1$
antiparallel neighbors, equals
\begin{eqnarray}
  \label{eq:fact}
  p(u_2\ |\ \sigma,u_1\to\sigma) &=& \frac{p_{\sigma,\sigma}(u_1,u_2)}
  {\sum_{u_2} p_{\sigma,\sigma}(u_1,u_2)} 
  \propto (K-u_2)\ p_{\sigma} (u_2) \ , \nonumber\\ 
  p(u_2\ |\ \sigma,u_1\to-\sigma) &=& \frac{p_{\sigma,-\sigma}(u_1,u_2)}
  {\sum_{u_2} p_{\sigma,-\sigma}(u_1,u_2)}
  \propto u_2\ p_{-\sigma} (u_2)\ ,
\end{eqnarray}
and it is thus independent on $u_1$, cf. the closure assumptions
(\ref{eq:independent}). 

Taking any microscopic configuration $\vec\sigma$, the distribution
$p_\sigma(u)$ is calculated as
\begin{equation}
  \label{eq:p_config}
  p_\sigma(u) = \frac 1N \sum_{i=1}^N \delta_{\sigma_i,\sigma}\ 
 \delta_{u_i(\vec \sigma), u} \ ,
\end{equation}
where $u_i(\vec \sigma) = \sum_j J_{ij} \delta_{\sigma_i,-\sigma_j}$
counts the number of unsatisfied links incident to $i$. Again,
in analogy to a microcanonical calculation, we should sum over all
configurations having exactly the same values of $p_\sigma(u)$ for all
$\sigma \in \{\pm 1\}$ and $u\in \{0,...,K\}$. As in the the previous
section, we can circumvent the explicite microcanonical calculation by
going to a generalized canonical ensemble by introducing formal
inverse temperatures $\beta_\sigma(u)$ for every pair $(\sigma,u)$.
These formal temperatures have to be adjusted in order to constrain the
$p_\sigma(u)$ to the desired values. Out of equilibrium, they
are not directly related to the physical temperature $\beta$.

We consequently have to determine the partition function
\begin{eqnarray}
  \label{eq:partition}
  Z &=& \sum_{\{\sigma_i\}_{i=1}^N} \exp\left\{ \sum_{\sigma,u}
    \beta_\sigma(u) \sum_{i=1}^N  \delta_{\sigma_i,\sigma}\ 
    \delta_{u_i(\vec \sigma), u} \right\} \nonumber\\
&=&  \sum_{\{\sigma_i\}_{i=1}^N} \exp\left\{ \sum_{i=1}^N
    \beta_{\sigma_i}(u_i(\vec \sigma)) \right\}\ .
\end{eqnarray}
Proceeding in close analogy with Sec. \ref{sec:model}, we introduce
partial partition functions $Z_{i|j}(\sigma_i, u_{i|j})$ for the
subtrees rooted in vertex $i$, with edges $(i,j)$ removed. The values
of $\sigma_i$ and $u_{i|j}$ are fixed, where $u_{i|j}$ denotes the
number of unsatisfied edges including vertex $i$ but not $j$.
Denoting the set of all neighbors of $i$ in the subtree by $V_{i|j}$,
the partition function can be calculated by an iterative procedure:
\begin{eqnarray}
  \label{eq:iteration_drt}
  Z_{i|j}(\sigma_i, u_{i|j}) = \sum_{I\subset V_{i|j}:\ |I|=u_{i|j}}
  && \prod_{k\in I} \left[ \sum_{u_{k|i}=0}^{K-1} Z_{k|i}(-\sigma_i,
    u_{k|i})\ \exp\left\{\beta_{-\sigma_i}(u_{k|i}+1)\right\} \right]
    \nonumber\\ 
  && \prod_{k\in V_{i|j}\setminus I} 
    \left[ \sum_{u_{k|i}=0}^{K-1} Z_{k|i}(\sigma_i,
    u_{k|i})\ \exp\left\{\beta_{\sigma_i}(u_{k|i})\right\} \right]\ .
\end{eqnarray}
Introducing generalized cavity fields as $h_{i|j}(\sigma,u)=\ln\{
Z_{i|j}(\sigma, u)/Z_{i|j}(-1,K-1) \}$, and looking for a homogeneous
solution $h_{i|j}(\sigma,u)=h(\sigma,u)$ for all edges $(i,j)$, we
obtain the following $2K-1$ self-consistency equations:
\begin{equation}
  \label{eq:self_cons}
  h(\sigma,u) = \ln \left\{ {K-1 \choose u}\ f(\sigma,-\sigma)^u\
    f(\sigma,\sigma)^{K-1-u} \right\}
 \end{equation}
with
\begin{equation}
  \label{eq:fss}
  f(\sigma,\tau) = \frac 
     {\sum_{u_1=0}^{K-1}
       e^{h(\tau,u_1)+\beta_{-\sigma}(u_1+\delta_{\sigma,-\tau})}} 
     {\sum_{u_1=0}^{K-1} e^{h(+,u_1)+\beta_{+}(u_1+1)}} \ .
\end{equation}
Note that, by plugging Eq. (\ref{eq:self_cons}) into (\ref{eq:fss}),
these equations can easily be reduced to only three self-consistent
equations for $f(+,+)$, $f(+,-)$ and $f(-,-)$, whereas $f(-,+)=1$ by
definition. Once these quantities are known, we can immediately
determine the distribution
\begin{equation}
  \label{eq:true_field}
  p_\sigma(u) = {K \choose u}\ f(\sigma,-\sigma)^u\
    f(\sigma,\sigma)^{K-u} e^{\beta_\sigma(u)}\ .
\end{equation}
As already said, the formal temperatures have to be adjusted 
in such a way that all
$p_\sigma(u)$ take the desired values.

\begin{figure}
  \begin{center}
    \includegraphics[angle=0,width=10cm]{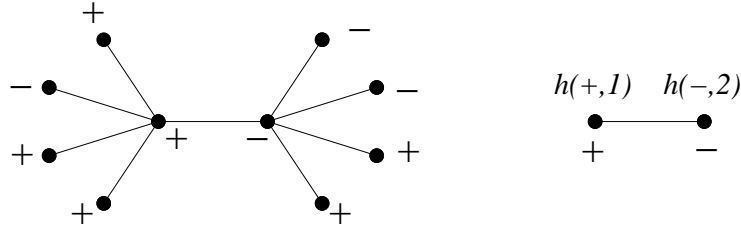}
    \caption{On the left, an example for a subgraph contributing
      to $p_{+-}(2,3)$ is given. Within the cavity calculation, the
      influence of the exterior part is replaced by the corresponding
      cavity fields $h(+,1)$ and $h(-,2)$, cf. Eq. (\ref{eq:joint}).}
    \label{fig:cavity_edge}
  \end{center}
\end{figure}
 
In order to show that the independent-neighbor approximation holds in
this ensemble, we have to calculate the joint distribution
$p_{\sigma_1,\sigma_2}(u_1,u_2)$ for neighboring vertices. As an
example we are analyzing the case $\sigma_1=+1$, $\sigma_2=-1$, the
other combinations work out analogously, see Fig.
\ref{fig:cavity_edge} for an illustration:
\begin{eqnarray}
  \label{eq:joint}
  p_{+-}(u_1,u_2) &\propto& e^{h(+,u_1-1)+h(-,u_2-1)+\beta_+(u_1)
    +\beta_-(u_2)} \\
  &\propto& {K-1 \choose u_1-1}\ f(+,-)^{u_1-1}\ f(+,+)^{K-u_1}
  {K-1 \choose u_2-1}\ f(-,-)^{K-u_2 }
  e^{\beta_+(u_1) +\beta_-(u_2)} \nonumber\\
  &\propto& u_1\ p_+(u_1)\ u_2\ p_-(u_2) f(+,-)^{-1}
\end{eqnarray}
For the conditional probability given in Eq. (\ref{eq:fact}) we
thus find
\begin{equation}
  \label{eq:conditional}
  p(u_2\ |\ +,u_1 \to -) = \frac{ p_{+-}(u_1,u_2)}{\sum_{u_2}
    p_{+-}(u_1,u_2)} \propto u_2\ p_-(u_2)
\end{equation}
which is exactly the independent neighbor approximation applied to
close the dynamical equations in Sec. \ref{sec:sites}.

This derivation can be easily generalized for the link approximation
scheme. We thus conclude that the presented approach is equivalent to
DRT under the assumption of replica symmetry. On the other hand,
focusing on finite-connectivity models makes this approach
more elegant and intuitively understandable, compared to fully
connected models for which the analogous equations are much more 
involved \cite{LaCoSh}.

\section{Conclusions and outlook}
\label{sec:conclusion}

To summarize, we have studied the dynamics of the Ising ferromagnet on a
Bethe lattice. This simple and, from the point of view of statics,
well-analyzed model serves as an ideal testing ground for a series of
dynamical approximation schemes first introduced in the context of the
analysis of stochastic local search optimization algorithms. In particular,
we have obtained a detailed characterization of the critical behavior of
this mean-field model.

Whereas the presented approximation schemes work very well in this simple
case, there remain crucial open questions in the
dynamic of disordered diluted systems, where the disorder can be
present either in fluctuating vertex degrees or in randomly
chosen interactions strengths. 

A first interesting application of our approach would therefore be the
study of disordered ferromagnets in their Griffith phase \cite{Br,BrHu}. 
This phase
is characterized by an anormally slow relaxation behavior even in the
high-temperature paramagnetic regime, 
which results from the existence of 
large regions of higher than average coupling strength.

An even more challenging problem appears if frustration is included
into the model, i.e. if we turn to systems displaying a low-temperature
spin-glass phase. Here the connection of our approach with DRT becomes
important and opens the way for the inclusion 
of replica symmetry breaking effects. A challenging task in this direction
would be to reproduce with such a dynamical approach the subtle phenomena
of cooling schedule dependence investigated numerically in \cite{MoRi-cooling}.

Finally, we want to mention as a future direction of research
the refined 
analysis of stochastic local search algorithms which solve (or approximate)
optimization problems like 3-satifiability or graph coloring. This
includes, e.g., the analysis of the influence of greedy heuristic
steps, which was out of range before \cite{PaWe}.

{\bf Acknowledgment:} We are very grateful to Leticia Cugliandolo,
R\'emi Monasson, Andrea Montanari, Andrea Pagnani, and Annette Zippelius 
for various interesting and
helpful discussions. We also thank the ICTP Trieste and the ISI
Foundation Turin for their hospitality during this work, and the
Exystence Network for financial support.

\end{document}